\documentstyle[aps,preprint]{revtex}
\begin{document}
\draft
\tightenlines
 
\title{Quasi-stationary distributions for stochastic processes with
an absorbing state}
\author{Ronald Dickman$^\dagger$ and Ronaldo Vidigal}
\address{
Departamento de F\'{\i}sica, ICEx,
Universidade Federal de Minas Gerais,\\
30123-970
Belo Horizonte - MG, Brasil\\
}

\date{\today}

\maketitle
\begin{abstract}
We study the long-time behavior of stochastic models with an absorbing 
state, conditioned on survival. 
For a large class of processes, in which saturation prevents
unlimited growth, statistical properties of the surviving sample
attain time-independent limiting values.
We may then define a {\it quasi-stationary} probability distribution 
as one in which the ratios $p_n(t)/p_m(t)$ (for any pair of 
nonabsorbing states $n$ and $m$), are time-independent.  
This is not a true stationary distribution, since the overall normalization
decays as probability flows irreversibly to the absorbing state.
We construct quasi-stationary solutions for the
contact process on a complete graph, the Malthus-Verhulst process,
Schl\"ogl's second model, and the voter model on a complete graph.  
We also construct the master equation and quasi-stationary state
in a two-site approximation for the contact process, and for
a pair of competing Malthus-Verhulst processes.

\end{abstract}

PACS: 05.10.Cg, 02.50.Ga, 05.40.-a, 05.70.Ln

Short title: Quasi-stationary distributions

\noindent {\small $^\dagger$electronic address: dickman@fisica.ufmg.br}

\newpage
\section{Introduction}

Stochastic processes with an absorbing state arise frequently in
statistical physics \cite{vankampen,gardiner}, epidemiology
\cite{bartlett}, and related fields.
On one hand we have autocatalytic processes (in lasers, chemical reactions, 
or surface catalysis, for example), and population models
(in genetic or epidemiological modeling), in which a population
of $n \geq 0$ individuals goes permanently extinct when the
absorbing state, $n\!=\!0$, is reached.  On the other hand,
phase transitions to an absorbing state in spatially extended
systems , exemplified by the contact process \cite{harris}, 
are currently of great interest
in connection with self-organized criticality \cite{socbjp}, 
the transition to turbulence \cite{bohr}, and, more generally, 
issues of universality in nonequilibrium
critical phenomena \cite{marro,hinrichsen}.  Such models are fequently 
studied using deterministic mean-field equations, Monte Carlo 
simulation, and renormalization group analyses.  
 
While fluctuations
play an essential role in the vicinity of the absorbing state, exact
solution of the master equation is not, in general, feasible.  A useful tool
in this situation is van Kampen's $\Omega$-expansion in inverse
powers of the system size, about the deterministic, macroscopic
solution.  It is clearly desirable to develop additional methods for 
analyzing
processes with absorbing states.  In this work we study models whose 
macroscopic equation admits a nonabsorbing (or {\it active}) stationary 
state, but which, for a
finite system size, must always end up in the absorbing state.
We show that the {\it quasi-stationary} distribution for such a system is 
readily constructed, and that it provides a wealth of information about its
behavior.   In fact, simulations of ``stationary" properties of lattice 
models with an absorbing state actually study the quasi-stationary regime,
since the only true stationary state (for a finite system) is the absorbing 
one.

For models without spatial structure, such as uniformly distributed 
populations, well stirred chemical reactors, or networks in which each 
node communicates equally
with all others, our results provide a complete description of
long-time properties, conditioned on survival.  In models with spatial 
structure, typified by nearest-neighbor interactions on a lattice, as in 
the contact process, the description in terms of one or a few variables 
corresponds to a mean-field theory that cannot adequately capture critical 
fluctuations.  Our approach does, nonetheless, allow one to put some of the 
fluctations ``back into" mean-field theory, and so we are
able to study finite size effects and moment ratios that are 
beyond the grasp of simpler approaches.  For one-step models involving a 
single variable, the computational demands of our approach are trivial; more 
complicated processes can be analyzed using straightforward numerical 
procedures.

The idea of the quasi-stationary state or distribution is quite simple.  
Consider a
Markov chain $n_t \geq 0$ in continuous time, with $n\!=\!0$ absorbing, and 
such that the macroscopic or rate-equation description (i.e., neglecting 
fluctuations) admits an active stationary state ($n_{st}>0$).  Then in
many cases one expects the probability distribution $p_n (t)$ to
bifurcate, after some initial transient, into two components, one with
all weight on the absorbing state, the other, $q_n (t)$, 
concentrated near the macroscopic value $n_{st}$, such that $p_n \approx 0$
for $n \simeq 0$.  The survival probability (for the process not to
have fallen into the absorbing state), is $P(t) = \sum_{n \geq 1} q_n(t)$.
We study the case in which $q_n$ has attained a time-independent form,
in which its only time dependence is through the overall factor $P_s(t)$.
Since the probability distribution is supposed to bifurcate, we may
write

\begin{equation}
p_n (t) = Q(t) \delta_{n,0} + q_n (t)
\end{equation}
where $Q(t) = 1-P(t)$, and by definition, $q_0 (t) = 0$.  
The quasi-stationary distribution (QS) is then 
defined via the condition:

\begin{equation}
p_n (t) = P(t) \overline{p}_n ,\;\;\;\; (n \geq 1),
\label{qshyp}
\end{equation}
where the $\overline{p}_n$ are time independent.  
(Again, $\overline{p}_0 \equiv 0$.) Since $P(t)$ is the
survival probability we have the normalization:

\begin{equation}
\sum_{n \geq 1} \overline{p}_n = 1.
\label{norm}
\end{equation}

Clearly, not every process with an 
absorbing state admits a QS distribution.  The
exponential decay process (with transition rates 
$W_{n,m} = m\delta_{n,m-1}$) and the birth-and-death process
($W_{n,m} = m\delta_{n,m-1} + m\lambda \delta_{n,m+1}$), for example,
do not.  While we will not try to determine rigorously the conditions
under which a QS distribution exists, it seems reasonable to suppose that
the macroscopic equation for the process in question admits an
active stationary solution.
All of the examples considered satisfy this condition.
The voter model represents a special case, in which the macroscopic
equation provides no information on the fate of the process, and the
QS distribution bears no resemblance to the stationary (absorbing)
distribution.

The balance of this paper is organized as follows.  In Sec. 2 we
illustrate the evaluation of the QS distribution, and the analysis of
associated statistical properties, for the simple case of the contact
process on a complete graph.  Sec. 3 is concerned with the closely
related Malthus-Verhulst process (MVP).  Both the contract process and 
the MVP exhibit a continuous phase transition in the infinite-size limit.
In Secs. 4 and 5 we study a models with a discontinuous phase transition:
the second Schl\"ogl process, and the voter model.  In Sec. 6 we return to 
the contact process, this time in a two-site approximation,
and in Sec. 7 study a pair of competing MVPs.  Sec. 8 contains a summary 
and discussion of our results.

\section{Contact process on a complete graph}

In the contact process (CP) \cite{harris,marro}, 
each site of a lattice is 
either occupied ($\sigma_i (t)= 1$),
or vacant ($\sigma_i (t)= 0$).  Transitions from $\sigma_i = 1$ to 
$\sigma_i = 0$ occur at a rate of unity, independent of the neighboring sites.  
The reverse can only occur if at least one neighbor is occupied: the 
transition from $\sigma_i = 0$ to $\sigma_i = 1$ occurs at a rate of 
$\lambda m$, where $m$ is the fraction of nearest neighbors of site $i$ 
that are occupied; thus the state $\sigma_i = 0$ for all $i$ is absorbing.  
The stationary order parameter $\rho$ (the fraction of occupied sites), is 
zero for $\lambda < \lambda_c$.  It is easy to show that in mean-field 
approximation (i.e., treating occupancy of sites as statistically 
independent events, and supposing that the density $\rho(t)$ is spatially 
uniform), the density follows the rate equation:

\begin{equation}
\frac{d \rho}{dt} = (\lambda -1) \rho - \lambda \rho^2 \;.
\label{cpmft}
\end{equation}
This equation predicts a continuous phase transition (from $\rho \equiv 0$ 
to $\rho = 1 -\lambda^{-1}$
in the stationary state) at $\lambda_c = 1$.  While this is qualitatively
correct, $\lambda_c $ is in general larger than 1; 
$\lambda_c = 3.29785(2)$ in one dimension, for example.

A mean-field description such as Eq. (\ref{cpmft}) does not, of course, yield
correct critical exponents for dimension $d < d_c = 4$, but that is not of 
concern here. Instead, we would like to study the simplest stochastic 
process for which Eq. (\ref{cpmft}) represents the macroscopic limit, 
(i.e., the limit of an infinite system, in which $\rho(t)$
is a deterministic variable).  By studying such a process we can recover the
fluctuations ignored in the macroscopic equation.  Since the process has an
absorbing state, its limiting ($t \to \infty$) distribution is trivial 
($p_n(t) \to \delta_{n,0}$), but we can study the long-time properties, 
conditioned on survival, using the QS distribution.

The stochastic birth-and-death process whose macroscopic limit is given
by Eq. (\ref{cpmft}) is the {\it contact process on a complete graph},  
in which the rate for a vacant site to turn occupied is $\lambda$ times the 
fraction of {\it all} sites that are occupied, rather than the fraction of 
nearest neighbors.  Since each site interacts equally with all others, 
all pairs of sites are neighbors,
defining a complete graph.  Given its connections with epidemic
modeling and percolation, this model has also been called ``percolitis" 
\cite{schulman}.

The state of the process is specified by a single variable $n$: the number of
occupied or infected sites.  This is a one-step process with nonzero 
transition rates

\begin{equation}
W_{n-1,n} = n
\end{equation}
\begin{equation}
W_{n+1,n} = \lambda \frac{n}{\Omega} (\Omega-n)
\end{equation}
on a graph of $\Omega$ sites.  These expressions yield the master equation

\begin{equation}
\frac{dp_n}{dt} = (n\!+\!1)p_{n+1} + 
(n\!-\!1)\overline{\lambda} (\Omega - n+1)p_{n-1} 
- \left[1 + \overline{\lambda} 
(\Omega-n) \right] n p_n \;,
\label{cpcgme}
\end{equation}
where $\overline{\lambda} \equiv \lambda/\Omega$.
Since $P(t) = \sum_{n \geq 1} p_n (t)$, we have $dP/dt = -p_1(t)$
in the contact process, and under the QS hypothesis, 
\begin{equation}
\frac{1}{P} \frac{dP}{dt} = - \overline{p}_1 .
\end{equation}

If we insert Eq. (\ref{qshyp}) in the master equation, then, dividing
through by $P(t)$ we obtain
\begin{equation}
\overline{\lambda} (n\!-\!1) [\Omega\!-\!n\!+1] \overline{p}_{n-1}
+(n\!+\!1)\overline{p}_{n+1} 
-[\overline{\lambda}(\Omega\!-\!n)+1]n \overline{p}_n 
+  \overline{p}_1 \overline{p}_n = 0 ,
\label{reccp}
\end{equation}
for $n \geq 1$.  For a given
$n$, Eq. (\ref{reccp}) furnishes a relation for $\overline{p}_{n+1}$ in
terms of $\overline{p}_{n}$ and $\overline{p}_{n-1}$.  Letting 
$u_n =  n[\overline{\lambda}(\Omega \!-\!n)+1]$,  we have

\begin{equation}
\overline{p}_{2}  = \frac{1}{2} \overline{p}_{1}(u_1 -\overline{p}_{1})
\end{equation}
and
\begin{equation}
\overline{p}_{n}  = \frac{1}{n} 
\left[(u_{n-1}-\overline{p}_{1})\overline{p}_{n-1}
+(n\!-\!2)(\Omega\!-\!2)\overline{\lambda} \overline{p}_{n-2} \right]
\end{equation}
for $n=2,...,\Omega$.
Thus the QS distribution is completely determined once $\overline{p}_{1}$
is known; the latter is determined via the normalization condition,
Eq. (\ref{norm}).

In practice the following iterative scheme converges very quickly.
Starting with a guess for $\overline{p}_{1}$ (unity, for instance),
one uses the recurrence relations to find the corresponding
$\overline{p}_{n}$, and then evaluates $S = \sum_{n= 1}^\Omega p_n $.
The procedure is then repeated using 
$\overline{p}_{1}' = \overline{p}_{1}/S$; after a modest number of steps,
$S$ converges to unity, at which point the $\overline{p}_{n}$ represent
the QS distribution.

We have confirmed that the master equation for the contact process
on a complete graph does in fact attain the QS distribution at long times.
We integrate the master equation using a fourth-order Runge-Kutta
scheme \cite{numrec}, and stop the integration when the mean population
conditioned on survival, $\langle n \rangle_s = \sum_n n p_n(t)/P(t)$,
changes by less than $10^{-10}$ per time step.  The resulting
distribution, $p^*_n = p_n(t)/P(t)$ is compared with the QS distribution
furnished via the recurrence relations in figure 1;  they are identical.

Having constructed the QS distribution and verified that it indeed
represents the long-time distribution (conditioned on survival), as given
by the master equation, we now examine some properties of the QS state.
Figure 2 shows the quasi-stationary population density $\rho = n/\Omega$
versus $\lambda$ for various system sizes.  This plot looks strikingly
similar to finite-size plots of the order parameter at a continuous
phase transition.  In
particular, it is evident that the $\lambda$-dependence is smooth
for any finite system, but that on increasing $\Omega$, the function
$\rho (\lambda,\Omega)$ approaches a singular limit.  

On finite-dimensional lattices the value of the
order parameter at the critical point is expected to scale as
$\rho(\lambda_c,\Omega) \sim \Omega^{-\beta/\nu_\perp} $, where $\beta$ and
$\nu_\perp$ are the critical exponents governing the order parameter
and the divergence of the correlation length \cite{marro}.
Here we find $\rho (\lambda_c,\Omega) \approx 0.685 \Omega^{-1/2}$ (from 
results for $\Omega=500 - 10^5$).
We note that the mean-field exponent
values $\beta=1$ and $\nu_\perp = 1/2$ for the contact process
would lead one to expect $\rho (\lambda_c,\Omega) \sim \Omega^{-2}$.
The reason for the difference would seem to be that on a complete graph
(where each site is, in effect, a unit distance from all others),
the correlation length is undefined, and so finite-size scaling ideas
do not apply.  
Instead, we can understand the exponent -1/2 by noting that the
microscopic equation (\ref{cpmft}), gives $\rho = 0$ at the critical
point, so that $n$ is purely a fluctuation.  The central limit theorem
then requires $n \sim \Omega^{1/2}$.
Away from the critical point, $\lambda = 1$, the finite-size
correction to the mean-field solution, $\rho = 1 - \lambda^{-1}$,
is ${\cal O} (\Omega^{-1})$.

Another property of interest at a continuous phase transition is
the moment ratio $m = \langle n^2 \rangle/\langle n \rangle ^2$.
This quantity is
analogous to Binder's reduced fourth cumulant 
\cite{binder}, at an equilibrium critical point: the curves 
$m(\lambda,\Omega)$ for various $\Omega$ cross near $\lambda_c$ (the 
crossings approach $\lambda_c$), so that $m$ assumes a universal
value $m_c$ at the critical point.  For the basic contact process
(and other models in the universality class of directed percolation),
$m_c = 1.174$ and 1.326 in one and two dimensions, respectively
\cite{rdjaff}.
In figure 3 we plot $m$ for the quasi-stationary contact process on a 
complete graph.  Qualitatively, the behavior is similar to that
observed on finite-dimensional lattices, with $m(\lambda)$ becoming
ever steeper with increasing system size; 
$|dm/d\lambda|_{\lambda_c} \sim \Omega^{1/2}$ for large $\Omega$.
The crossings approach $\lambda_c$, as expected.
We find that $m(1,\Omega) \simeq 1.660 + {\cal O} (\Omega^{-1/2})$ 
for large $\Omega$.

As noted above, the survival probability decays as $dP(t)/dt = -p_1 (t)$.
We may therefore interpret $\overline{p}_1$ as the quasi-stationary
decay rate.  The QS lifetime, $\tau = 1/\overline{p}_1$, grows  
as $\exp [\mbox{const.} \; \Omega (\lambda - \lambda_c)]$ for 
$\lambda > \lambda_c$,
as can be seen from figure 4, where we plot $\Omega^{-1} \ln \tau $ versus
$\lambda$ for several system sizes.  At the critical point, 
$\tau \sim \Omega^{1/2}$.

It may appear surprising that we are able to discuss quasi-stationary
properties for $\lambda < \lambda_c$, where an active stationary state 
does not exist, even in the thermodynamic limit.  The existence of
nontrivial QS properties for $\lambda < \lambda_c$ is in fact a
finite-size effect, analogous to a nonzero magnetization in the Ising
model above the critical temperature, on a finite lattice.  
For $\lambda > \lambda_c$ the QS state converges, as
$\Omega \to \infty$, to the true stationary state, while for 
$\lambda < \lambda_c$ it converges to the absorbing state.  But since the
properties of any {\it finite} system are nonsingular, we must expect
a smooth decay of, for example, the density, as $\lambda \to 0$.

\subsection{Pseudo-stationary distribution}

The contact process on a finite graph (be it a regular
lattice, a complete graph, or small-world network) has only the
absorbing state as its true stationary state.  It is nonetheless
instructive to examine the result of {\it forcing} a time-independent
solution on the master equation.  This is easily done by introducing the
generating function $F(z,t) \equiv \sum_n p_n(t) z^n$.  
The master equation, Eq.(\ref{cpcgme}), is equivalent to the 
partial differential equation,

\begin{equation}
\frac{\partial F}{\partial t} = (1-z)\left[
(1-\overline{\lambda} z) \frac{\partial F}{\partial z} 
+ \overline{\lambda} z \frac{\partial }{\partial z}
\left( z \frac{\partial F}{\partial z} \right) \right] \;.
\label{gfpss}
\end{equation}
Setting $\partial F/\partial t = 0$ and letting 
$G(z) = \partial F/\partial z$, a first integration
yields

\begin{equation}
G(z) = C z^{\Omega-1} e^{1/\overline{\lambda} z} \;,
\label{gofz}
\end{equation}
where $C$ is a constant.  Integrating once more gives

\begin{equation}
F_{ps}(z) = C \sum_{n=0}^{\infty} 
\frac{z^{\Omega-n}}{\overline{\lambda}^n n! (\Omega-n)} \;.
\label{fpss}
\end{equation}
While the corresponding distribution,

\begin{equation}
p_{ps,m}  = \frac{C \overline{\lambda}^m}{(\Omega-m)! m}
\label{ppss}
\end{equation}
yields $d p_m/dt = 0$ 
when inserted in the master equation, it is not a valid
solution since it cannot be normalized, and does not satisfy the
correct boundary condition at $m=0$.  For this reason we
call $p_{ps,m}$ a {\it pseudo-stationary} distribution.
Similar pseudo-stationary distributions have been derived by Mu\~noz
from the Fokker-Planck equation for continuous versions of the contact
process and of a model with multiplicative noise \cite{munoz98}.

If we treat $p_{ps,m}$ as a proper 
probability distribution, restricting it to $m=1,...,\Omega$, and
normalizing it on this set,  then the resulting distribution
turns out to be very close to the quasi-stationary one
when $p_m $ is negligible in the vicinity of $m=0$.
Figure 5 (for $\Omega=100$) shows that the two distributions are 
essentially identical for $\lambda=2$ (as they are for larger
$\lambda$ values), and for $\lambda=1.5$, except for small $n$.
For $\lambda=1.2$, however, the distributions are radically
different; in particular, we see that the pseudo-stationary
distribution does not respect the absorbing boundary at $n=0$.

In summary, the pseudo-stationary distribution provides a simple and
useful approximation to the true QS distribution for $\lambda$ 
(and $\Omega$) sufficiently large that $p_m \simeq 0$ for $m \simeq 1$.
(In this case the survival probability decays extremely slowly,
rendering the hypothesis of a stationary distribution - while
strictly incorrect - at least reasonable.)
As we approach the critical value, however, the region near
$n=0$ assumes dominance, and $p_{ps}$
bears little relation to the quasi-stationary distribution.

\section{Malthus-Verhulst process}

The Malthus-Verhulst process (MVP) is a birth-and-death type process,
$n(t)$, very similar to the contact process on a complete graph,
but with the difference that $n$ may assume any nonnegative
integer value.  The nonzero transition rates for this one-step 
process are:

\begin{equation}
W_{n-1,n} = \mu n + \frac{\nu}{\Omega} n(n\!-\!1) \;,
\end{equation}
\begin{equation}
W_{n+1,n} = \lambda n \;.
\end{equation}
In the first expression, $\Omega$ represents the system size,
as discussed by van Kampen \cite{vankampen}.  In the MVP there is no
lattice structure, and no
fixed limit to the population size, as there is in the contact process;
rather, growth is limited by the term $\propto \nu$, which represents
competition between individuals for access to finite resources.
(Thus $\nu$ represents an intrinsic competition parameter, and
$ (\nu/\Omega)n(n\!-\!1)$ is the rate of interactions between pairs of
individuals in a system of size $\Omega$.)  The saturation effect
that permits a nontrivial quasi-stationary state is imposed in the
death term, whereas it appears in the birth term, in the contact process.

Letting $n = \Omega \rho$, the macroscopic equation for the MVP
is

\begin{equation}
\frac{d \rho}{dt} = (\lambda \!-\! \mu) \rho - \nu \rho^2
\label{mamvp}
\end{equation}
which has the stationary solution $\overline{\rho} = 
(\lambda \!-\! \mu)/\nu$.  The master equation is

\begin{equation}
\frac{dp_n}{dt} = [\mu+\overline{\nu} n](n\!+\!1)p_{n+1} + 
(n\!-\!1)\lambda p_{n-1} 
- [\lambda + \mu +\overline{\nu}(n\!-\!1)] n p_n \;,
\label{mvpme}
\end{equation}
with $\overline{\nu}= \nu/\Omega$.  Inserting the quasistationary
form of the probability distribution, one readily obtains
the recurrence relation

\begin{equation}
\overline{p}_n = \frac{(q_{n-1} - \mu\overline{p}_1) \overline{p}_{n-1}
       -\lambda(n-2) \overline{p}_{n-2}}{n[\mu + \overline{\nu}(n\!-\!1)]} ,
\label{mvprr}
\end{equation}
where
\begin{equation}
q_n = [\lambda + \mu +\overline{\nu}(n\!-\!1)] n
\end{equation}
The QS distribution can be found using the same iterative procedure
as for the contact process.  (We cut off the distribution at a certain
$M$ such that $\overline {p}_ \leq 10^{-20}$ for $ n > M$.)

The quasi-stationary properties of the MVP are quite similar to 
those of the contact process on a complete graph.  Figure 6 shows the
dependence of $\rho$ on $\lambda$ for system sizes $\Omega = 100$, 
10$^3$ and 10$^4$.  (We fix $\mu = \nu = 1$ in all numerical studies  
of the MVP.)   The density grows linearly with $\lambda$
for $\lambda > \lambda_c = \mu$, unlike in the contact process, where
the macroscopic saturation term is also proportional to the birth rate,
$\lambda$.  As in the contact process on a complete graph, the lifetime 
$\tau \sim \Omega^{1/2}$ at $\lambda_c$, and the QS density
decays $\sim \Omega^{-1/2}$. 

The moment ratios $m$ (figure 6, inset) show the same
qualitative trend as in the CP.  The asymptotic value of $m$ at the
critical point, moreover, is the same as in the CP: we find
$m \simeq 1.660 - 0.814 \Omega^{-1/2}$ for large $\Omega$.
This identity of $m$ values suggests that the limiting QS probability 
distributions at $\lambda_c$ are the same for the two processes.  This 
is verified numerically; for $\Omega = 10^3$, for example, the distributions 
differ by less than about $2 \times 10^{-4}$.

In fact, the critical
QS probability distribution for the MVP (and by extension for the
contact process) enjoys the scaling property

\begin{equation}
\overline{p}_n = \frac{1}{\sqrt{\Omega}} f (n/\sqrt{\Omega}) \;.
\label{scqsc}
\end{equation}
The scaling function $f$ may be found using a method that parallels
van Kampen's analysis of the master equation \cite{vankampen}.  First, we 
treat $n$ in Eq. (\ref{mvprr}) as a continuous variable, expanding 
$\overline{p}_{n \pm 1}$ to second order.  Setting $\lambda = \mu$ at the 
critical point, we obtain

\begin{equation}
\frac{1}{2} [(2\mu + \overline{\nu})n +  \overline{\nu}n^2] 
\frac{d^2 \overline{p}}{dn^2}
+ [2 \mu +  \overline{\nu}n(1+n)] \frac{d \overline{p}}{dn}
+ 2 \overline{\nu}n \overline{p} = -\mu \overline{p}_1 \overline{p} \;.
\label{rrcont}
\end{equation}
Next we make the change of variables $y = n/\Omega^{1/2}$, and 
$f(y) = \Omega^{1/2} \overline{p}_n$.  This yields an expansion in inverse 
powers of  $\Omega^{1/2}$; at lowest order ($\Omega^{-1}$) we find

\begin{equation}
\mu y \frac{d^2 f}{dy^2} + [2 \mu + \nu y^2] \frac{df}{dy} 
+ 2 \nu y f = - \mu f(0) f(y) \;.
\label{def}
\end{equation}
The initial value $f(0)$ must be chosen such that $\int_0^\infty f dy = 1$.

To integrate Eq. (\ref{def}) numerically we let $g(y) = df/dy$, and note that
$g(0) = -[f(0)]^2/2$, and $dg/dy|_{y=0} = -(2\nu/\mu) f(0)$.
Setting $\mu = \nu = 1$, numerical integration yields the scaling function
shown in figure 7, which agrees well with the distributions for the MVP
and the contact process on a complete graph, in the limit of large
system size.  
[Note that the change of variables used to derive Eq. (\ref{def}) yields
the very same equation when applied to the recursion relation for
the contact process, Eq. (\ref{reccp}), at its critical point, 
$\lambda \!=\! 1$.]
The asymptotic scaling function $f(y)$ yields the
moment ratio $m=1.660063$, again in agreement with results from
the recurrence relations.  (Note that $m = f(0)/\langle y \rangle$, as 
may be verified directly from Eq. (\ref{def}).)

Despite some superficial differences, the MVP and the contact process
on a complete graph show the same type of phase transition (continuous,
with mean-field-like critical exponents), in the limit $\Omega \to \infty$,
and the same scaling properties at
their respective critical points.  

As in the case of the contact process, one may define a pseudo-stationary 
distribution by demanding that the r.h.s. of the master equation be
zero.  The resulting expression,

\begin{equation}
p_{ps,n} = C\frac{(\lambda/\overline{\nu})^n}
	{n (n + \mu/\overline{\nu} -1)!}
\label{pssmv}
\end{equation}
is very similar to that of the contact process on a complete graph, and
again represents a good approximation to the quasi-stationary
distribution if $\overline{p}_n \simeq 0$ for $n \simeq 1$.

\section{Schl\"ogl's Second Model}

The processes studied in the preceding sections exhibit a continuous
phase transition; next we consider a process with a discontinuous
transition.  Schl\"ogl's second model \cite{schlogl} describes the
set of autocatalytic reactions $A \to 0$, $2A \to 3A$, and $3A \to 2A$
in a well stirred system.  Since the growth process is quadratic
in the density (rather than linear, as in the contact and MV processes),
a low-density active state is unstable, and we expect a discontinuous
transition to a state with a nonzero density as the growth rate
is increased.  The nonvanishing transition rates are:

\begin{equation}
W_{n-1,n} = \mu n + \frac{\nu}{\Omega^2} n(n\!-\!1)(n\!-\!2) \;,
\end{equation}
\begin{equation}
W_{n+1,n} = \frac{\lambda}{\Omega} n (n\!-\!1)\;.
\end{equation}
These yield the macroscopic equation:

\begin{equation}
\frac{d\rho}{dt} = - \mu \rho + \lambda \rho^2 - \nu \rho^3  \;,
\label{macssp}
\end{equation}
which admits an active stationary solution,

\begin{equation}
\rho_s = \frac{1}{2}\left[ \lambda + \sqrt{\lambda^2 - 4 \mu \nu} \right] ,
\label{rsssp}
\end{equation}
for $\lambda \geq 2 \sqrt{\mu \nu}$  (The stationary density
jumps from zero to $\sqrt{\mu \nu}$ at the transition.  
It is known that in a spatially extended system the transition is
actually continuous for $d < 4$ \cite{grass82,brachet}, but here
we treat the well stirred system, which exhibits a discontinuous 
phase transition.)

Since this is a one-step process, it is a simple matter to derive
the recurrence relations for the quasi-stationary probabilities: 

\begin{equation}
\overline{p}_n = \frac{(q_{n-1} - \mu \overline{p}_1)\overline{p}_{n-1}
- \overline{\lambda} (n\!-\!2)(n\!-\!3)\overline{p}_{n-2}}
{n[\mu + \overline{\nu}(n\!-\!1)(n\!-\!2)]}
\label{rrssp}
\end{equation}
where
\begin{equation}
q_n = [\mu + \overline{\lambda} (n\!-\!1)
      + \overline{\nu}(n\!-\!1)(n\!-\!2)]n \;,
\end{equation}
with $\overline{\lambda} = \lambda/\Omega$ and
$ \overline{\nu} = \nu/\Omega^2$.
These may be solved using the same method as for the contact process
or the MVP.

In the following numerical example, we fix $\mu = \nu = 1$, and study 
the neighborhood of the transition at $\lambda=2 $.  Figure 8 shows      
the quasi-stationary density near the transition; with increasing
$\Omega$ the density approaches the discontinuous stationary solution,
Eq. (\ref{rsssp}).  
The density goes from ${\cal O}(1/\Omega)$
to ${\cal O}(1)$, in a transition region of width $\sim \Omega^{-1/2}$.
The moment ratio $m$, at the transition, appears to approach unity 
(from above) slowly, as $\Omega \to \infty$; data for 
$\Omega \leq 5 \times 10^4$ yield $m \simeq 1 + 2.64 \Omega^{0.265}$.
Similarly, the density at the transition approaches its limiting value
from below: $\rho \simeq 1 - 1.5 \Omega^{0.211}$.  The lifetime
at the transition appears to grows as 
$ \tau \simeq \Omega^{0.35}$.  Essentially the same power laws (but 
with different prefactors) are found at the transition for
$\mu=2$ and $\nu=1/2$.

The probability distribution is bimodal in the neighborhood of the
transition, a hallmark of a discontinuous transition.  This is
evident in figure 9, where the relative amplitudes of the two
peaks, one at $n=1$, the other near $n \simeq 1000$, shift rapidly
near the transition.  The bimodal distribution of course presages
a hysteresis loop when $\Omega \to \infty$.

\section{Voter model on a complete graph}

A somewhat simpler model exhibiting a discontinuous phase transition
is the voter model \cite{liggett,chate}.  
It differs from the systems considered previously in that it possesses
{\it two} absorbing states.
In the voter model each site on a lattice
is in one of two states, $A$ or $B$; each $A-B$ nearest-neighbor pair
has a unit rate to change to the state $AA$ (with probability $\omega$)
or to $BB$ (with probability $1-\omega$).  Let $n(t)$ denote the
number of sites in state $A$; both $n=0$ and $n=\Omega$
are absorbing, on a lattice of $\Omega$ sites.  Starting from a
random initial condition with equal densities of $A$ and $B$, the
stationary probability distribution switches discontinuously from
$\delta_{n,0}$ to $\delta_{n,\Omega}$ as $\omega$ is increased
through 1/2.  Of particular interest is the coarsening dynamics
for $\omega = 1/2$ \cite{chate}.  
In this case $\langle n \rangle$
is constant, and the probabilities of the two absorbing states
are determined by $n(0)$.

In this brief section we study 
the model on a complete graph, for the case $\omega = 1/2$.
The master equation is

\begin{equation}
\frac{dp_n}{dt} = \frac{n\!-\!1}{2\Omega}(\Omega\!-\!n\!+\!1)p_{n-1} + 
  \frac{n\!+\!1}{2\Omega}(\Omega\!-\!n\!-\!1)p_{n+1}
-  \frac{n}{2\Omega}(\Omega\!-\!n)p_n  \;.
\label{mvvote}
\end{equation}
This yields the macroscopic equation $d\rho/dt = 0$, reflecting the
dominance of fluctuations in this process.
From the terms for $n=0$ and $n=\Omega$ we find that in the
QS state, the survival probability obeys

\begin{equation}
\frac{1}{P}\frac{dP}{dt} = - \frac{1}{2} (1\!-\!\Omega^{-1})
    (\overline{p}_1 \!+\! \overline{p}_\Omega) \;,
\end{equation}
allowing us to write the following relation for the QS distribution:

\begin{equation}
- \overline{p}_n (1\!-\!\Omega^{-1})
(\overline{p}_1 \!+\! \overline{p}_{\Omega-1})
=  \frac{n\!-\!1}{\Omega}(\Omega\!-\!n\!+\!1) \overline{p}_{n-1}
 +  \frac{n\!+\!1}{\Omega}(\Omega\!-\!n\!-\!1) \overline{p}_{n+1}
 -  2 \frac{n}{\Omega}(\Omega\!-\!n\!) \overline{p}_n \;.
\end{equation}
The solution is the uniform distribution, $\overline{p}_n =1/(\Omega-1)$
for $n=1,...,\Omega\!-\!1$ (here, $\overline{p}_0 = 
\overline{p}_{\Omega} = 0$).

In the case of the voter model, the QS distribution looks very different
from the stationary distribution, even in the infinite-$\Omega$
limit.  This is because the process has no active stationary state:
on a complete graph it must always fluctuate into one of the
absorbing states.  
If we exclude all trials that hit one of
the boundaries, we are in effect looking at a random walker confined
to an interval, whose distribution is asymptotically uniform.

\section{Two-site approximation for the contact process}

In this section we return to the contact process, constructing a
stochastic description in terms of nearest-neighbor pairs of sites,
on a ring of $\Omega$ sites.  The idea of the approximation
is similar to that of two-site or pair mean-field dynamic mean-field
theory \cite{marro}, except that in the present case we treat the 
number of occupied sites, $n$, and the number of doubly-occupied
nearest-neighbor pairs, $z$, as stochastic variables.  When
necessary, we invoke the usual factorization of three-site probabilities
in terms of those for two sites, in order to construct the transition rates.

We begin by establishing the range of allowed values for $z$.
Using `1' and `0' to represent occupied and vacant sites, respectively, 
$z$ is the number of (11) nearest-neighbor (NN) pairs.  Let $w$ represent the
number of (10) NN pairs.  (For obvious reasons, the number of  (01) pairs is
again $w$.)  Note that $w$ is not an independent variable, since
each 1 is followed wither by a 0 or another 1, yielding $n = z + w$.
Similarly, the number of (00) pairs $v$ is given by $v = \Omega - 2n + z$.
(Here we used $n + 2w + z = \Omega$.)
Evidently $v$ is restricted to the set $\{0,...,\Omega\}$ while
$w$ must be in $\{0,...,\Omega/2\}$ (We assume $\Omega$ even.)
These considerations imply certain limits for $z$ on a ring of
$\Omega$ sites, as listed in table 1.

\begin{center}
{\bf Table 1.} Allowed values for $z$ in the CP on a ring.
\begin{tabular}{|c|c|} 
\hline\hline
$n$ &$z$ \\
\hline\hline
0, 1    &  0     \\
\hline
$2,..., \Omega/2$    & $ 0,...,n\!-\!1$   \\
\hline
$ \Omega/2 +1,..., \Omega\!-\!1$  & $2n\!-\!\Omega,...,n-1   $    \\
\hline
$ \Omega $   & $\Omega$    \\
\hline\hline
\end{tabular}
\end{center}

Next we construct the transition rates $W_{n',z';n,z}$.  From a given
state $(n,z)$ there are at most five possible transitions, viz., to the
states $(n',z') = (n\!-\!1,z)$, $(n\!-\!1,z\!-\!1)$, $(n\!-\!1,z\!-\!2)$, 
$(n\!+\!1,z\!+\!1)$, and $(n\!+\!1,z\!+\!2)$.  The first three
represent annihilation of a particle with, resp. zero, one or two
occupied nearest neighbors, while the last two represent creation
of a particle at a site with one, or two occupied neighbors, resp.
The rate of the transition $(n,z) \to (n\!-\!1,z)$ is the number of
isolated particles, that is, triplets of the form (010).  This
number is not determined completely by $n$ and $z$;
we estimate it as the number of (01) pairs times the 
conditional probability of a vacant site given an occupied neighbor:
$w \times (w/n)$.  This estimate, however, is obviously wrong when $z=n-1$
and $n \geq 2$, in which case there are {\it no} (010) triplets.  Thus
we have

\begin{equation}
W_{n\!-\!1, z;n, z} = 
\left\{ \begin{array} {ll}
 0 , & z=n\!-\!1,\;n>1\\
\frac{(n\!-\!z)^2}{n} , & \mbox{ otherwise } 
	\end{array} \right.
\label{w1}
\end{equation}
By similar arguments, we find:

\begin{eqnarray}
W_{n\!-\!1, z\!-\!1;n, z} &=& 
\left\{ \begin{array} {ll}
 2 , & z=n\!-\!1, \; n>1\\
(n\!-\!z)\frac{2z}{n} , & \mbox{ otherwise }  \\
	\end{array} \right.
\\
W_{n\!-\!1, z\!-\!2; n, z} &=& 
\left\{ \begin{array} {ll}
 0 , &  z=1\\
 n\!-\!2 , & z=n\!-\!1, \;n = 3,...,\Omega-1\\
\frac{z^2}{n} , & \mbox{ otherwise } 
	\end{array} \right.
\\
W_{n\!+\!1, z\!+\!1; n, z} &=& 
\left\{ \begin{array} {ll}
 \lambda , & z=n\!-\!1 \\
 \lambda (n\!-\!z)\frac{\Omega \!-\!2n\!+\!z}{\Omega \!-\!n} , 
 & \mbox{ otherwise } 
	\end{array} \right.
\\
W_{n\!+\!1, z\!+\!2; n, z} &= &
\left\{ \begin{array} {ll}
 0 , & z=n\!-\!1\\
\lambda \frac{(n\!-\!z)^2}{\Omega\!-\!n} ,& \mbox{ otherwise } 
	\end{array} \right.
\end{eqnarray}

In the macroscopic limit, the master equation defined by these rates yields 
a pair of coupled equations for the particle density $\rho$ and the pair 
density $\zeta = z/\Omega$ \cite{marro}.  The critical value 
$\lambda_c\! =\! 2$ in this approximation.

Since this is not a one-step process, 
we cannot obtain the QS distribution via recurrence relations, as 
was done in the previous examples.
The most obvious approach - direct integration of the master equation -
works reasonably well for modest system sizes ($\Omega < 200$ or so), but 
becomes very slow for large $\Omega$.  This is due in part to the large 
number of  equations ($ \sim \Omega^2$), and to the fact that
the transition rates are ${\cal O} (\Omega)$, forcing one to use a time step 
$ \sim \Omega^{-1}$, to avoid numerical instability.  A much more efficient 
iterative scheme that converges directly to the QS distribution is described 
in Ref. \cite{intme}. The basic idea of this approach is that in the 
QS regime, 

\begin{equation}
\overline{p}_n = \frac{r_n}{w_n - r_0} \;, 
\label{qssit}
\end{equation}
for any nonabsorbing state $n$, where $r_n = \sum_{n'} W_{n,n'} p_{n'}(t) $ 
is the flux of probability into state $n$ ($r_0$ is the flux into the 
absorbing state), and $w_n = \sum_{n'} W_{n',n} $ is the total
transition rate exiting $n$.  The iterative scheme \cite{intme} 
starts with an arbitrary distribution $p_n$ normalized on $n \geq 1$ 
(or, in the multivariate case, on the set of nonabsorbing states).  Next 
one evaluates

\begin{equation}
p'_n = a p_n + (1\!-\!a) \frac{r_n}{w_n - r_0}  \;,
\label{itqss}
\end{equation}
where $a$ is a parameter between zero and one.  The new distribution 
$p'_n$ is then normalized, and inserted in the r.h.s. of 
Eq. (\ref{itqss}), and the process repeated
until it converges.  For large $\Omega$ the time to reach the QS 
distribution is three
or more orders of magnitude smaller than that required for numerical 
integration.
(In the present case convergence is enhanced if one uses a small value 
for $a$, e.g., 0.01.  A further economy is realized by noting that, 
near $\lambda_c$, $\overline{p}_{nz} $ is extremely small for 
$n \simeq \Omega/2$, allowing one to
truncate the distribution.  For $\Omega = 1280$ and $\lambda = 2$, 
for example, $\overline{p}_{nz} \sim 10^{-17}$ for $n=350$.)

We studied the QS properties of the contact process on a ring of 
$\Omega = 10$, 20, 40,...,
2560 sites, using the pair approximation.   Figure 10 shows that the 
dependence of the particle density and the moment ratio
on $\lambda$ and on system size is qualitatively similar to that found
in the complete graph case.  The scaling laws for the density 
($\propto \Omega^{-1/2}$) and the lifetime ($\propto \Omega^{1/2}$)
at the critical point are the same as for the CP on a complete graph.
The critical values of the moment ratio $m$ at $\lambda_c$, extrapolated
to infinite system size, yield $m_c = 1.653$, nearly the same as for the
CP on a complete graph.  The critical probability distribution 
$\overline{p_n}$ again shows a scaling
collapse (figure 11), but the scaling function differs somewhat
from that for the complete graph.  Figure 12 shows the joint QS distribution
$\overline{p}_{n,z}$ at the critical point, for $\Omega = 40$.  For each $n$
there is a most probable number, $z^*(n)$, of pairs; $z^*$ grows
linearly with $n$.

\section{Competing Malthus-Verhulst processes}

As a final example we consider a population consisting of two
types, $A$ and $B$, that evolve according to the MVP
analyzed in Sec. 3.  (The two subpopulations may, for example,
possess different alleles of a certain gene, in a species with asexual 
reproduction.)  $A$ and $B$ interact via two mechanisms: (1) each individual 
competes with all others (regardless of type) for access to resources; 
(2) on reproducing, $A$ mutates to $B$ with probability $q$, and vice-versa.
Thus we have the following set of transition rates:

\begin{equation}
W_{n_A-1, n_B; n_A ,n_B} = \mu_A n_A 
+ \frac{\nu}{\Omega} n_A (n_A \!+\!n_B \!-\!1) \;,
\end{equation}
\begin{equation}
W_{n_A, n_B-1;n_A, n_B} = \mu_B n_B 
+ \frac{\nu}{\Omega} n_B (n_A \!+\! n_B \!-\!1) \;,
\end{equation}
\begin{equation}
W_{n_A+1, n_B; n_A,n_B} = \lambda_A p n_A + \lambda_B q n_B \;,
\end{equation}
\begin{equation}
W_{n_A ,n_B+1; n_A,n_B} = \lambda_B p n_B + \lambda_A q n_A\;,
\end{equation}
where $p=1\!-\!q$.

Evidently the state $n_A \!=\! n_B \!=\! 0$ is absorbing.  For
mutation probability 
$q=0$, $n_A \!=\! 0$ is also absorbing (similarly $n_B \!=\! 0$),
and in the QS state only one type is present, i.e.,
$\overline{p}_{n_A, n_B} = 0$ if both $n_A$ and $n_B$ are nonzero.
This prompts us to investigate the effect of a 
nonzero mutation rate:
Will the predominance
of one species persist when mutation is possible?  To address this
issue we study the order parameter

\begin{equation}
\Delta \equiv \frac{\langle |n_A \!-\! n_B| \rangle}
{\langle n_A + n_B \rangle}
\label{Delta}
\end{equation}
in the QS state.  Figure 13 shows how $\Delta$ varies with $q$, for
various system sizes.  The competing MVPs are both 
at the critical point, $\lambda_i = \mu_i$.  (For simplicity we take
all rates, $\lambda_i$, $\mu_i$, and $\nu$, equal to unity).
The order parameter decreases with increasing mutation probability, 
as expected, and with increasing system size.  From analysis of
$\Delta$ at various $q$-values, as a function of $\Omega$, it appears
that $\lim_{\Omega \to \infty} \Delta (q,\Omega;\lambda_c) = 0$ for {\it any}
nonzero $q$.  In other words, an arbitrarily small mutation rate 
restores the symmetry of the process.
We observe qualitatively similar behavior in $\Delta (q,\Omega)$
above the critical point ($\lambda > 1$).
At the critical point, the decay of $\Delta (q,\Omega)$ with $\Omega$ 
is slower than exponential,
but faster than a power law; it can be fit approximately by  
a stretched exponential $\sim \exp[- c\; \Omega^{0.2}]$, where the factor
$c$ depends on $q$ (see figure 13 inset).

\section{Discussion}

We study quasi-stationary probability distributions
for Markov processes exhibing a phase transition between an active and an
absorbing state.  The QS distribution, which represents the long-time
limit of the probability distribution, conditioned on survival, yields a
complete statistical description of the process in this limit.
While the QS distribution can always, in principle, be found by
integrating the master equation, we present much more efficient
methods for its generation: recurrence relations for one-step
processes, and an iterative scheme for multi-step or multivariate
processes.

Some insight into the nature of the quasi-stationary distribution 
is afforded by the considering the equation
(in the notation of Sec. VI), for $n \geq 1$,

\begin{equation}
\frac{dq_n}{dt} = -w_n q_n + r_n + r_0 q_n \;,
\label{qme}
\end{equation}
where $r_n = \sum_{n'} W_{n,n'} q_{n'}$ is the flux of probability 
into state $n \geq 0$.  (As usual we take state 0 as absorbing.)
Without the final term on the right-hand side, this is simply the
master equation.  Including this term, however, the equation has
as its stationary solution the QS distribution,
$\overline{p}_n$.  In other words, the QS distribution corresponds
to the stationary state of a ``process" in which all flux to the
absorbing state is immediately {\it reinserted} into the non-absorbing 
subspace.  The portion alloted to state $n$ is equal to its 
probability, $q_n$.  

Thus the QS distribution does not result from
imposing a simple boundary condition (such as periodic, or
reflection at the origin), on the original master equation.
Since Eq. (\ref{qme}) is nonlinear, it is not a master equation. 
It may happen, nevertheless, that the
QS distribution for a certain process (possessing an absorbing state),
is also the {\it stationary} distribution for some other process,
as in fact was found for the voter model.

In this work we study QS distributions for processes with an absorbing state,
but without spatial structure.  We find that the Malthus-Verhulst
process and the contact process on a complete graph have the same
scaling properties near their respective
critical points.  A study of Schl\"ogl's second model illustruates
how a discontinuous phase transition emerges in the infinite-size limit,
associated with a rapid change in the QS distribution, which is bimodal
in the vicinity of the transition.  
The voter model on a complete graph serves as a negative example: 
it has no active stationary state, and the QS distribution, which
is uniform on the set of allowed density values, is very different
from the true long-time distribution, even in the infinite-size limit.
Our two final examples are bivariate
processes.  The pair approximation to the contact process shows 
scaling properties that are very similar to those of the complete-graph
version. A study of a pair of competing Malthus-Verhulst processes reveals 
that mutation causes a radical change in the QS distribution.

Several directions for future work on quasi-stationary distributions
can be mentioned.  First, it should be possible to derive continuum
equations, analogous to the Fokker-Planck equation, to describe
the QS distribution in the limit of large system size.  Second,
analysis of QS properties for a series of system sizes promises
to be a useful method for studying lattice models possessing an
absorbing state.  Through finite-size scaling analyses and extrapolation
procedures, it should be possible to extract useful estimates
of critical properties for such models.  Finally, numerical study
of more elaborate models used in epidemiology and population dynamics
may yield a better understanding of such processes.
\vspace{1em}

\noindent{\bf Acknowledgment}

We are grateful to Miguel Angel Mu\~noz for helpful discussions.
This work is supported by CNPq, Brazil.

\newpage
\noindent FIGURE CAPTIONS

\noindent {\bf Figure 1.} Probability distributions for the contact process 
on a complete graph, $L=100$ sites.  The curves (connecting values at integer 
points) represent the long-time limit of the master equation, normalized to
the survival probability $P(t)$.  Points represent the quasi-stationary
distribution determined by the recurrence relations.  Left curve: 
$\lambda = 1.2$; right: $\lambda = 2$.
\vspace{1em}

\noindent {\bf Figure 2.} Quasi-stationary population density versus 
$\lambda$ for the contact process on a complete graph.  System sizes 
100, 200, 500, 1000, and 2000, top to bottom on the left-hand side.  The 
inset is a detail of the critical region, including the curve for the 
infinite-size limit, $\rho - 1 - \lambda^{-1}$.
\vspace{1em}

\noindent {\bf Figure 3.} Quasi-stationary moment ratio $m$ versus 
$\lambda$ for the contact process on a complete graph.  System sizes 
20, 50, 100, 200, 500, and 1000, in order of increasing steepness.
\vspace{1em}

\noindent {\bf Figure 4.} Quasi-stationary lifetime $\tau$ versus $\lambda$ 
for the contact process on a complete graph.  System sizes 100, 500, 
1000, and 5000, upper to lower on left-hand side.
\vspace{1em}

\noindent {\bf Figure 5.} Quasi-stationary distribution (solid line) and 
pseudo-stationary distribution (points) for the contact process on a 
complete graph, $L=100$ sites for $\lambda = 2$ (upper), 1.5 (middle),  
and 1.2 (lower).  The inset in the middle panel is a detail of the
region near $n=0$.
\vspace{1em}

\noindent {\bf Figure 6.} Quasi-stationary density versus $\lambda$ for the
Malthus-Verhulst process.  System sizes 100, 10$^3$, and 
10$^4$, upper to lower on the left-hand side.  The inset shows
moment ratio $m$ versus $\lambda$ for the same system sizes.
\vspace{1em}

\noindent {\bf Figure 7.} Scaling plot of the quasi-stationary distribution,
$f = \Omega^{1/2} \overline{p}_n$ versus $y = n/\Omega^{1/2}$,
at the critical point: $+$: MVP, $\Omega = 10^4$;
$\circ$: contact process on complete graph, $\Omega = 10^3$; solid curve: 
asymptotic scaling function $f(y)$
obtained via numerical integration.
\vspace{1em}

\noindent {\bf Figure 8.} Quasi-stationary density versus $\lambda$ 
for the second Schl\"ogl process with $\mu=\nu=1$.  
System sizes 100, 500, 1000, and 
5000, (in order of increasing steepness).  
The dashed curve is the mean-field solution, which is discontinuous 
at $\lambda = 2$.
\vspace{1em}

\noindent {\bf Figure 9.} Quasi-stationary distribution for the
second Schl\"ogl process with $\mu=\nu=1$, system size $\Omega = 1000$.  
$\lambda = 1.98$, 2, and 2.02 (upper to lower on left-hand side).
\vspace{1em}

\noindent {\bf Figure 10.} Quasi-stationary density versus $\lambda$ for the
contact process in the pair approximation.  System sizes 20, 40, 80, 
160, 320, and 640.  The inset shows
moment ratio $m$ versus $\lambda$ for the same system sizes.
\vspace{1em}

\noindent {\bf Figure 11.} Scaling plot of the quasi-stationary distribution:
$P^* = \Omega^{1/2} \overline{p}_n$ versus $n^* = n/\Omega^{1/2}$,
at the critical point of the contact process. Open circles: pair
approximation, $\Omega = 640$;
solid line: pair approximation, $\Omega = 2560$;
filled circles: contact process on complete graph, $\Omega=1000$.
\vspace{1em}

\noindent {\bf Figure 12.} Quasi-stationary probability distribution for the 
critical contact process in the pair approximation, system size $\Omega = 40$.  
\vspace{1em}

\noindent {\bf Figure 13.} Quasi-stationary order parameter $\Delta$ versus 
mutation rate $q$ for a pair of competing MVPs at the critical point.
System sizes $\Omega$ = 20, 50, 100, 200, 500, 1000 (upper to lower).
The inset shows $\ln \Delta$ versus $\Omega^{0.2}$ for (upper to lower)
$q=0.02$, 0.05, and 0.1.
\vspace{1em}


\begin{thebibliography}{99}

\bibitem{vankampen}
	  van Kampen N G 1992 
	  {\it Stochastic Processes in Physics and Chemistry}
	  (Amsterdam: North-Holland)

\bibitem{gardiner}
	  Gardiner C W 1990 
	  {\it Handbook of Stochastic Methods}
	  (Berlin: Springer-Verlag)

\bibitem{bartlett} 
	  Bartlett M S 1960
	  {\it Stochastic Population Models 
	  in Ecology and Epidemiology} 
	  (London: Methuen)

\bibitem{harris} 
	  Harris T E 1974 
	  {\it Ann. Prob.} {\bf 2} 969
 
\bibitem{socbjp} 
	  Dickman R, Mu\~noz M A, Vespignani A and Zapperi S 2000
	  {\it Braz. J. Phys.} {\bf 30} 27 

\bibitem{bohr}
	  Bohr T, van Hecke M, Mikkelsen R and Ipsen M 2001
	  {\it Phys. Rev. Lett.} {\bf 86} 5482, and references therein.

\bibitem{marro} 
	  Marro J and Dickman R 1999
	  {\it Nonequilibrium Phase Transitions in Lattice Models} 
	  (Cambridge: Cambridge University Press)

\bibitem{hinrichsen}
	  Hinrichsen H 2000
	  {\it Adv. Phys.} {\bf 49} 815

\bibitem{schulman}
	  Schulman L S and Seiden P E 1986
	  {\it Science} {\bf 233} 425

\bibitem{numrec} 
	  Press W H, Flannery B P, Teukolsky S A and
	  Vetterling W T 1996
	  {\it Numerical Recipes}
	  (Cambridge: Cambridge University Press)

\bibitem{binder}
	  Binder K 1981 
	  {\it Phys. Rev. Lett.} {\bf 47} 693

\bibitem{rdjaff} 
	  Dickman R and Kamphorst Leal da Silva J (1998)
	  {\it Phys. Rev.} E{\bf 58} 4266

\bibitem{munoz98}
	  Mu\~noz M A 1998
	  {\it Phys. Rev.} E{\bf 57} 1377

\bibitem{schlogl}
	  Schl\"ogl F 1972
	  {\it Z. Phys.} {\bf 253} 147

\bibitem{brachet}
	  Brachet M E and Tirapegui E 1981
	  {\it Phys. Lett.} A{\bf 81} 211

\bibitem{grass82}
	  Grassberger P 1982
	  {\it Z. Phys.} B{\bf 47} 465

\bibitem{liggett} 
	  Liggett T 1985 
	  {\it Interacting Particle Systems} 
	  (Berlin: Springer-Verlag)

\bibitem{chate}
	  Dornic I, Chat\'e H, Chave J and Hinrichsen H 2001
	  {\it Phys. Rev. Lett.} {\bf 87} 045701

\bibitem{intme}
	  Dickman R 2001 
	  {\it Preprint} cond-mat/0110xxxx.

\end{thebibliography}
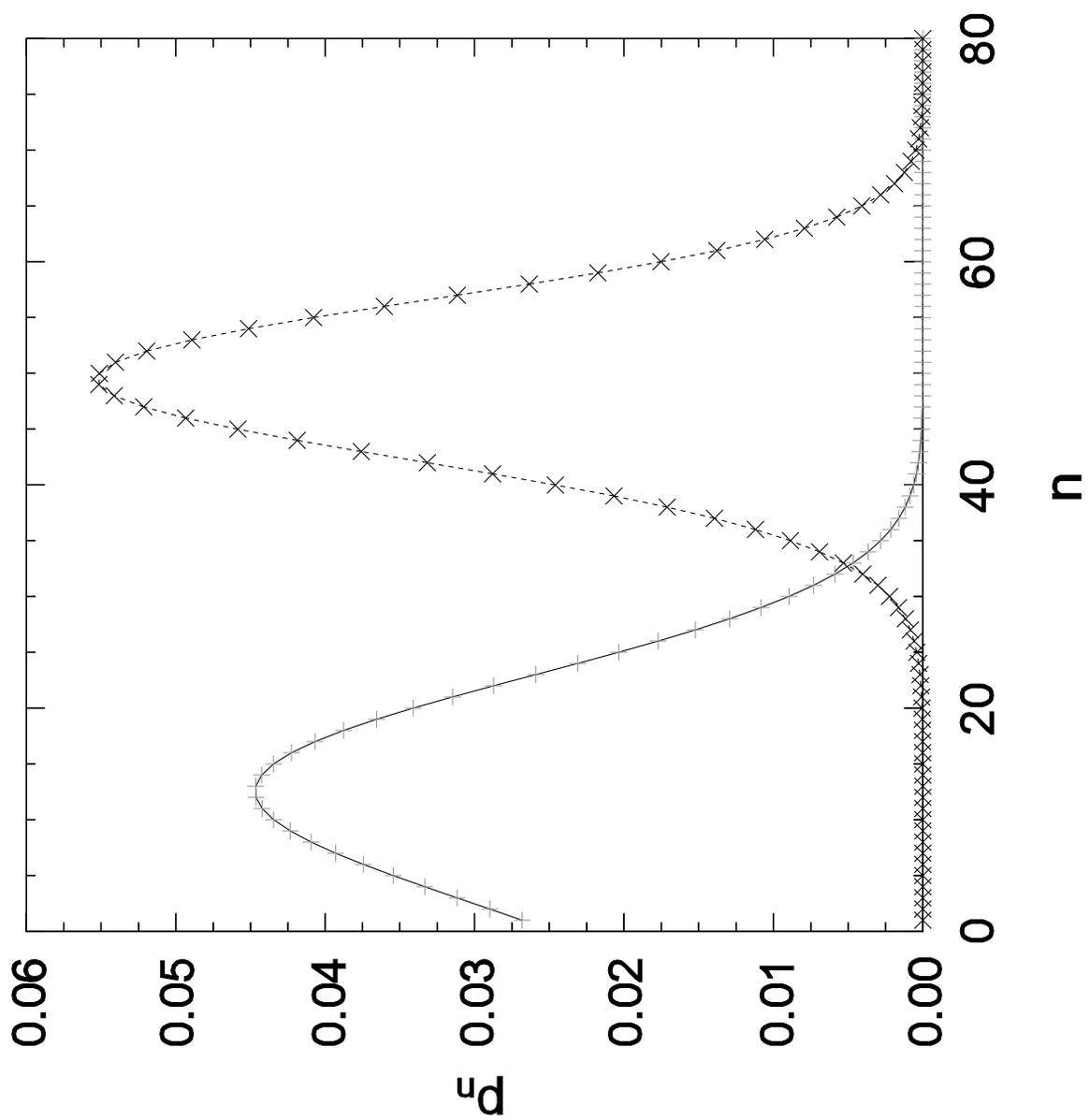
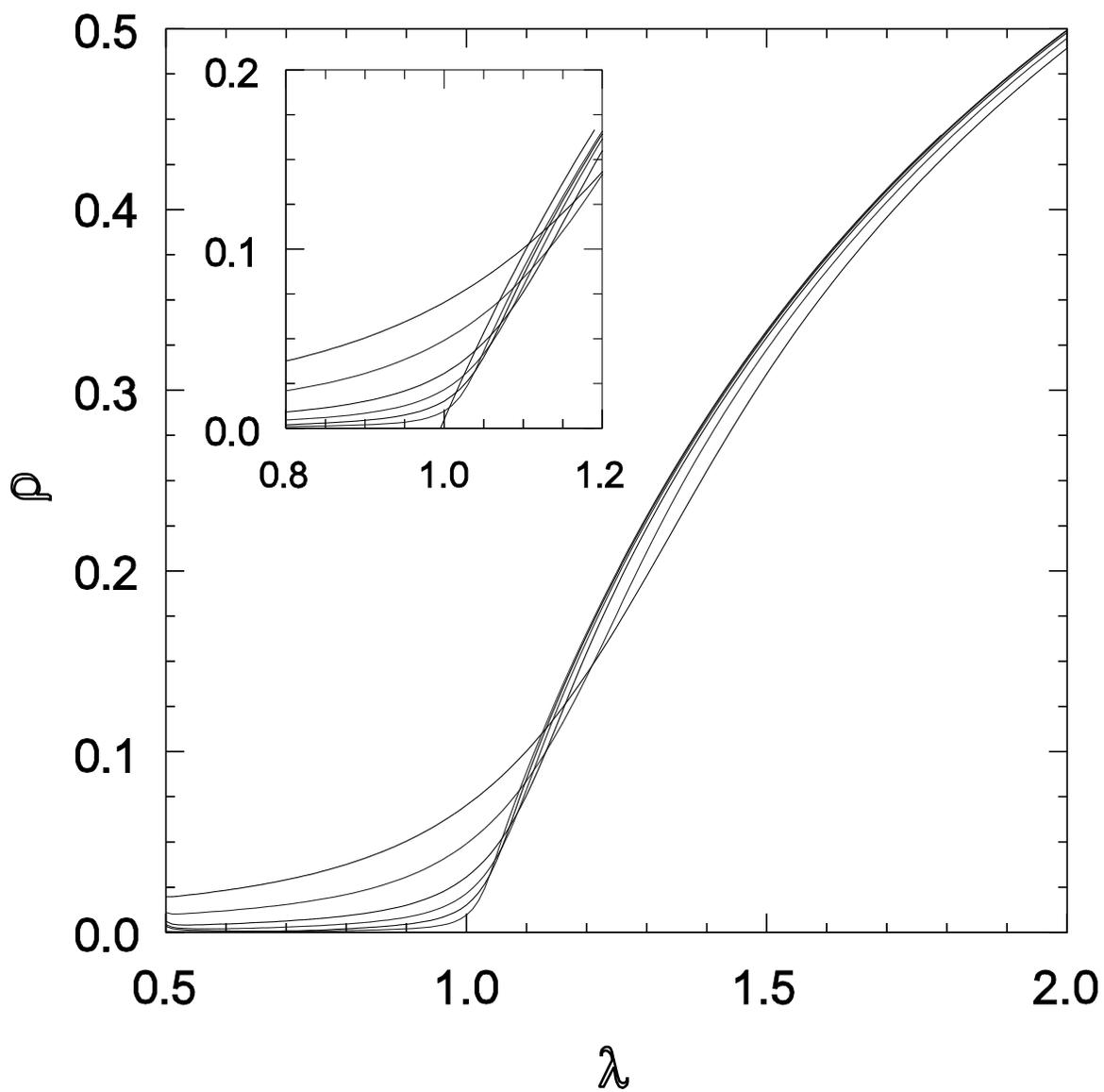
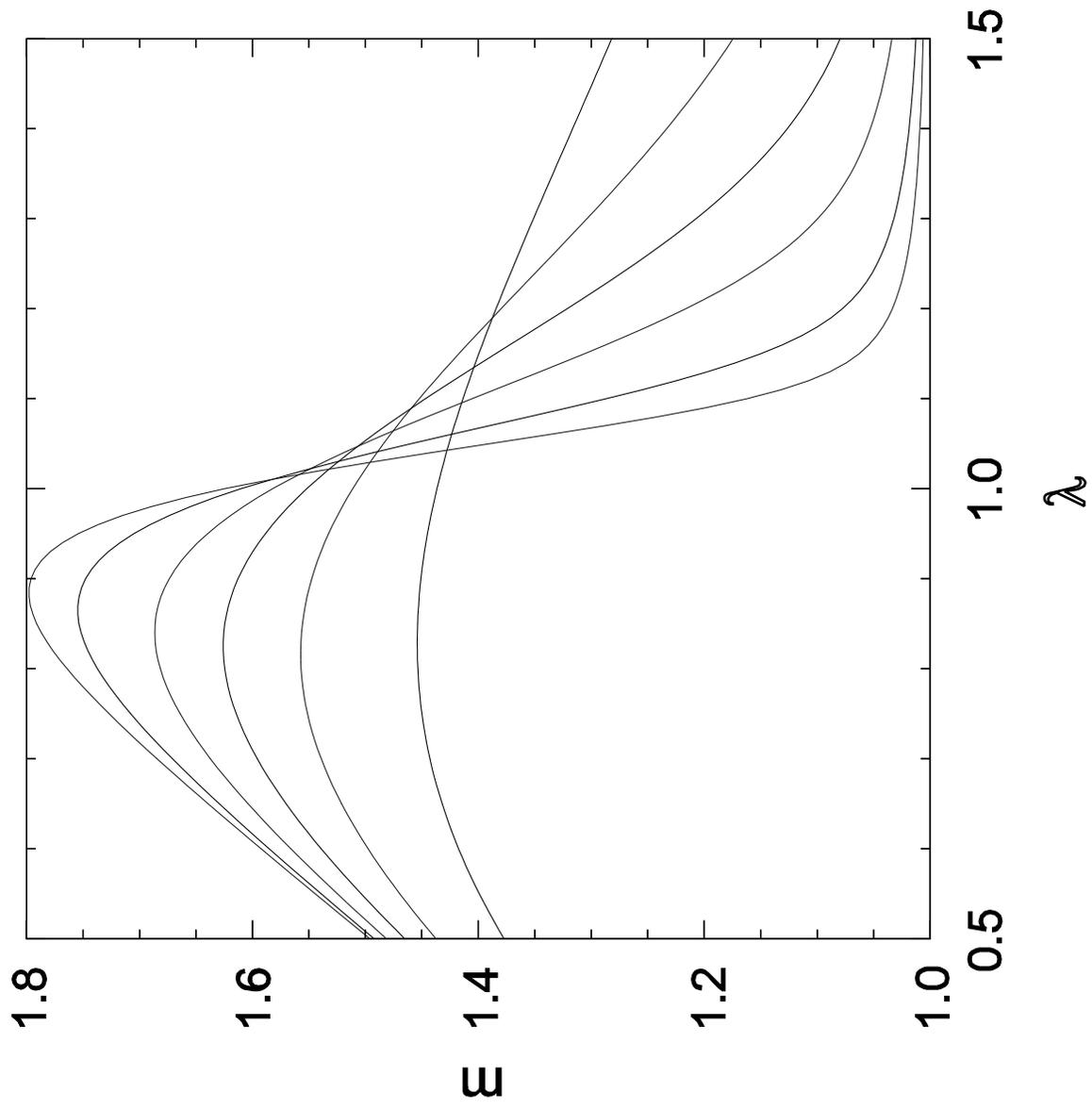
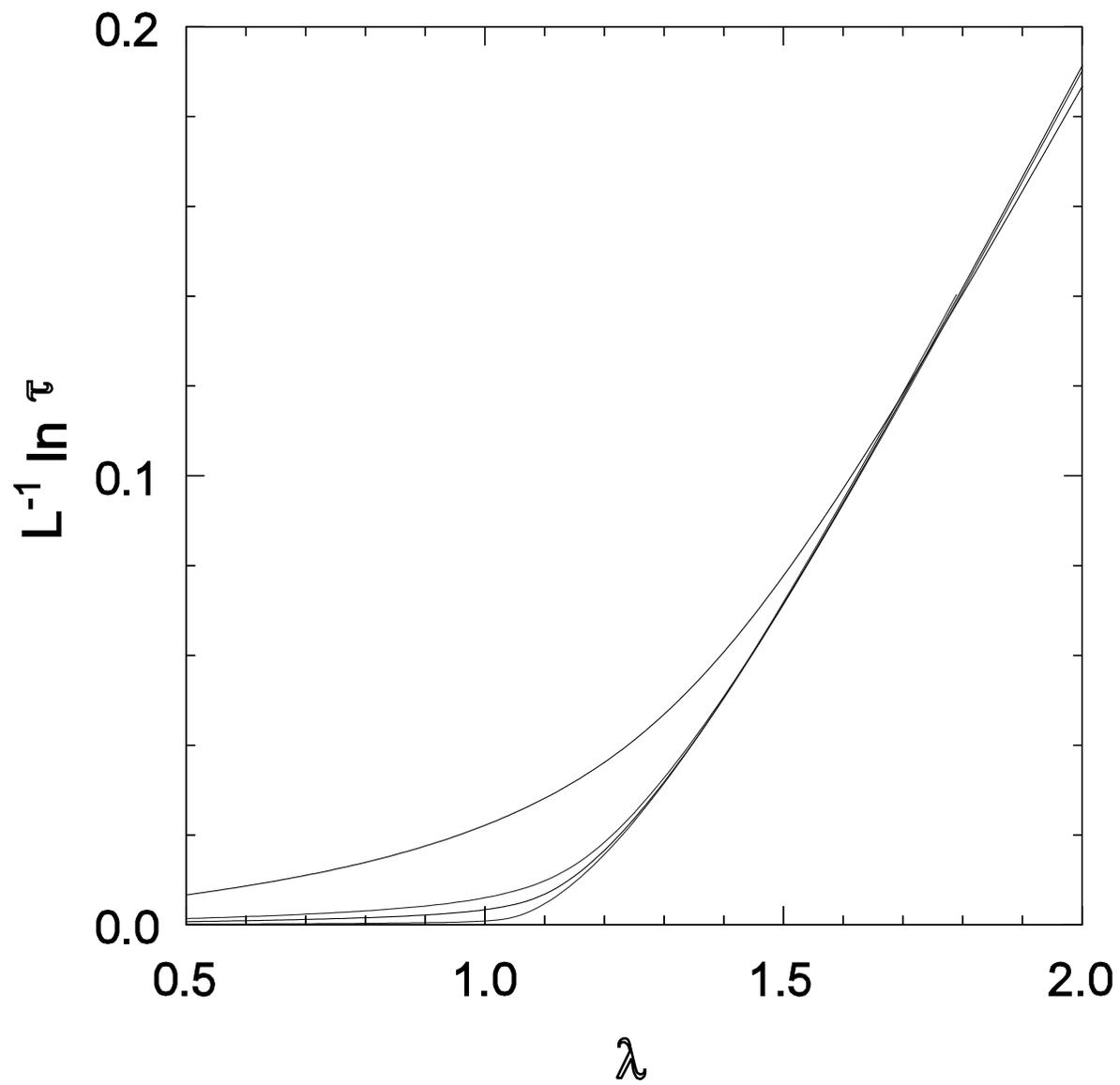
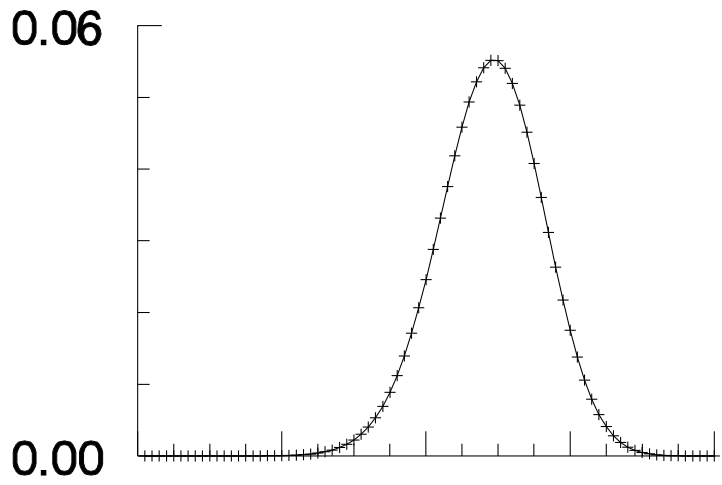
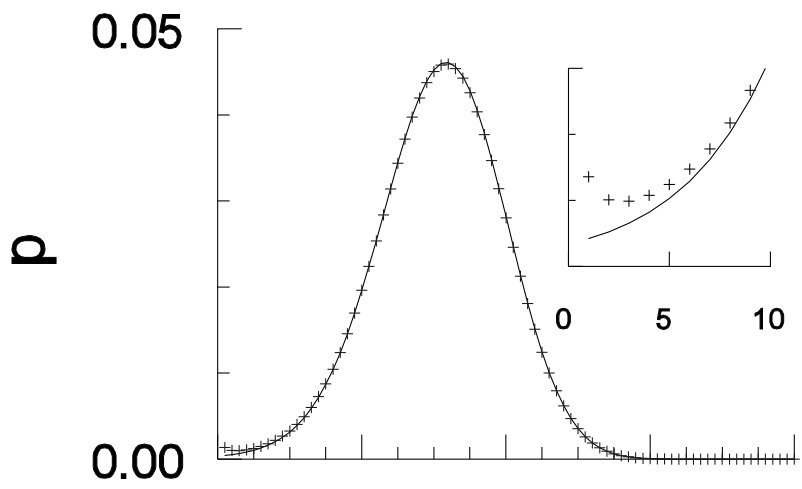
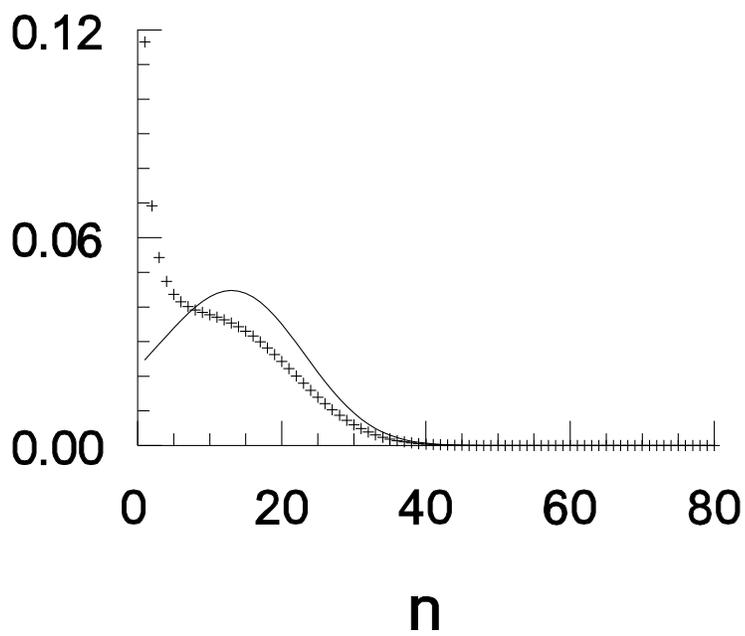
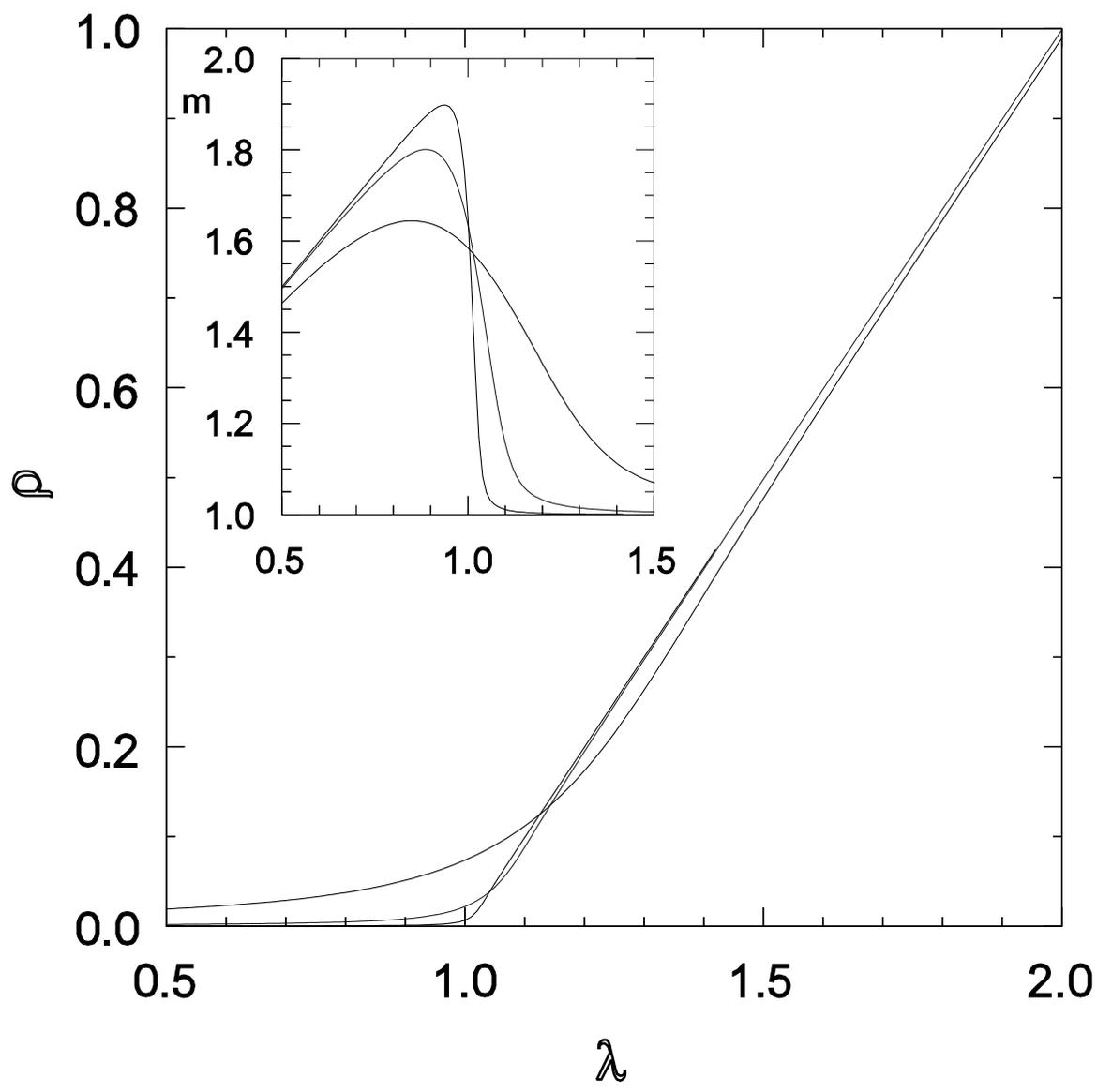
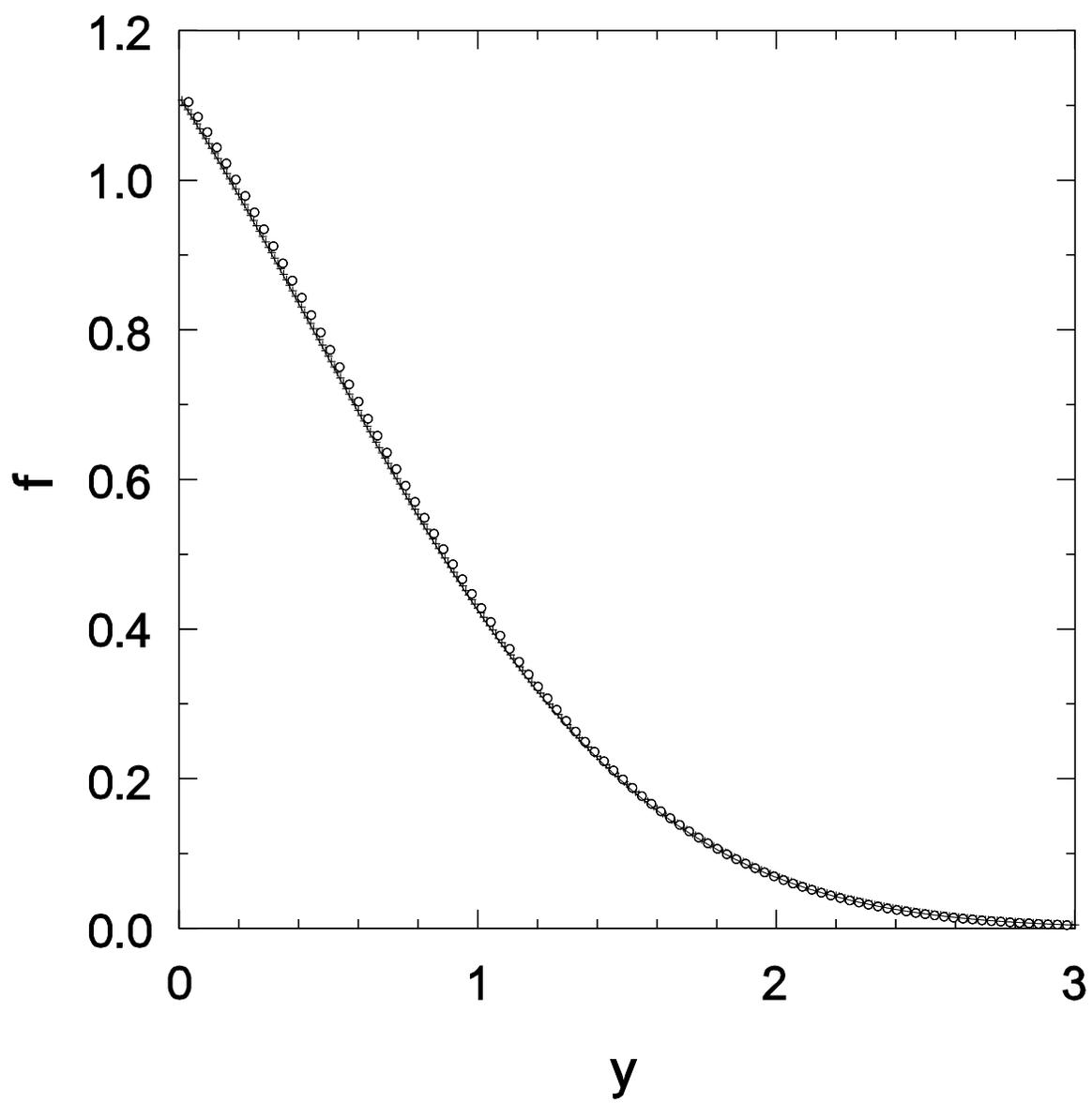
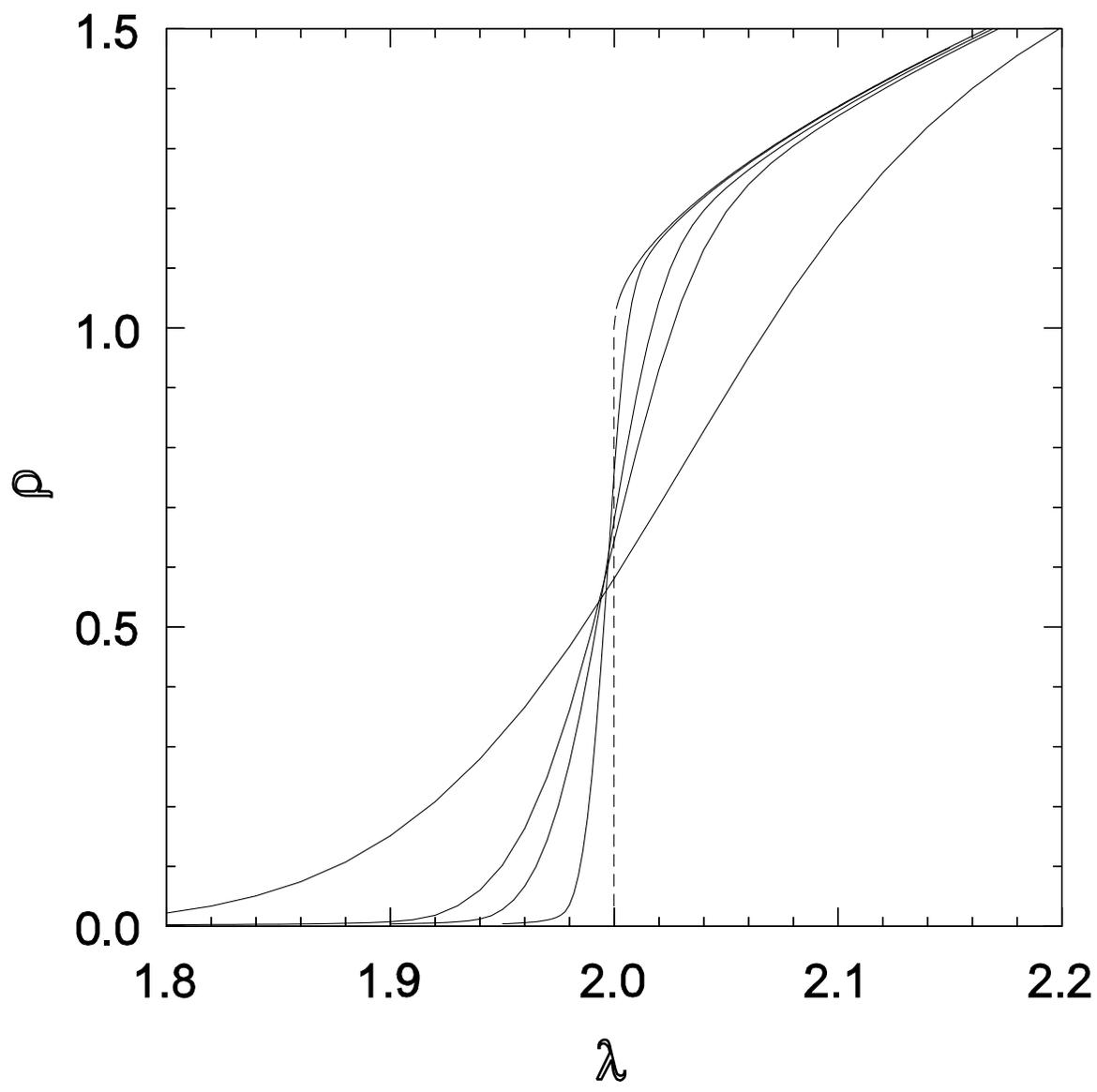
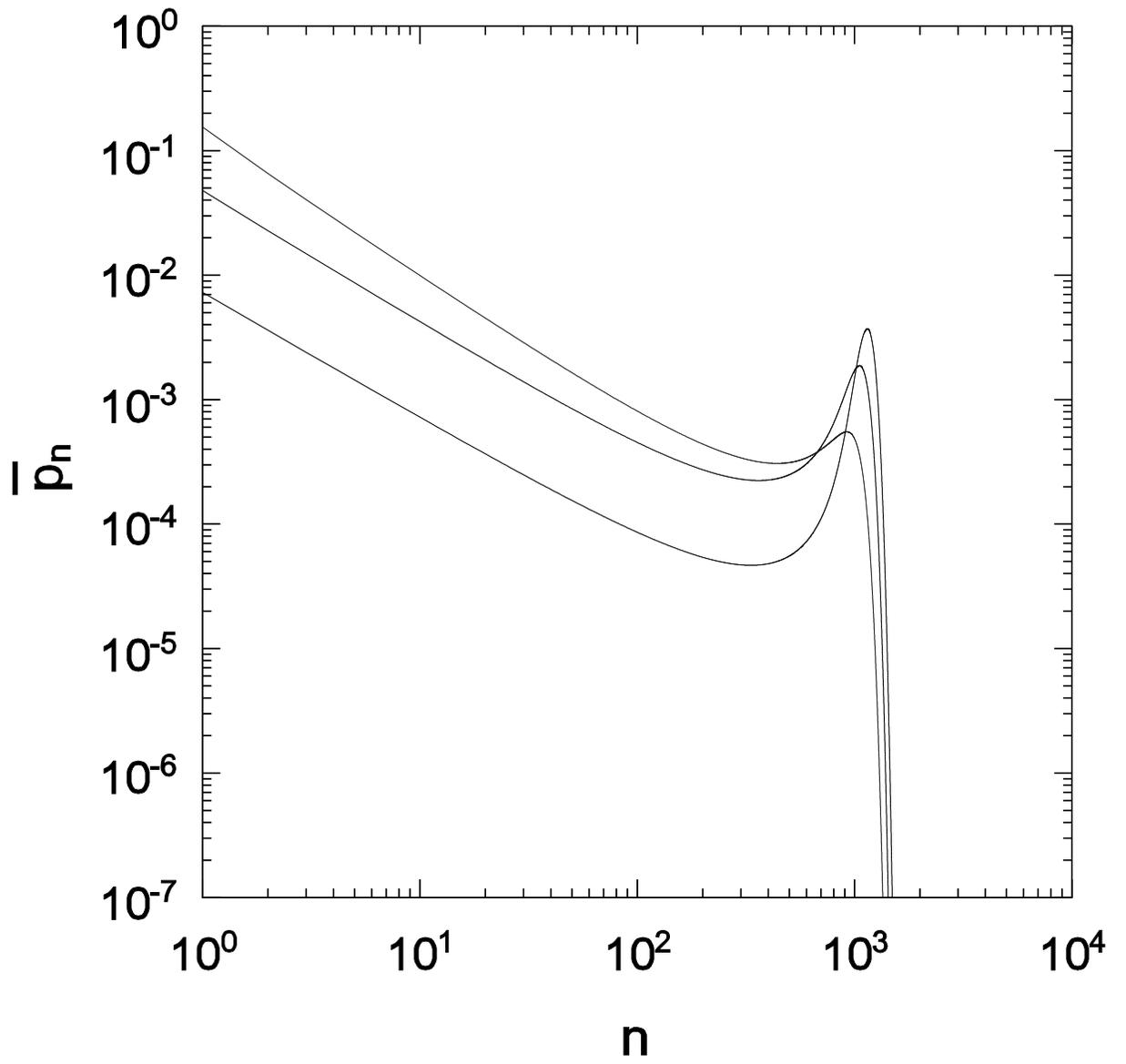
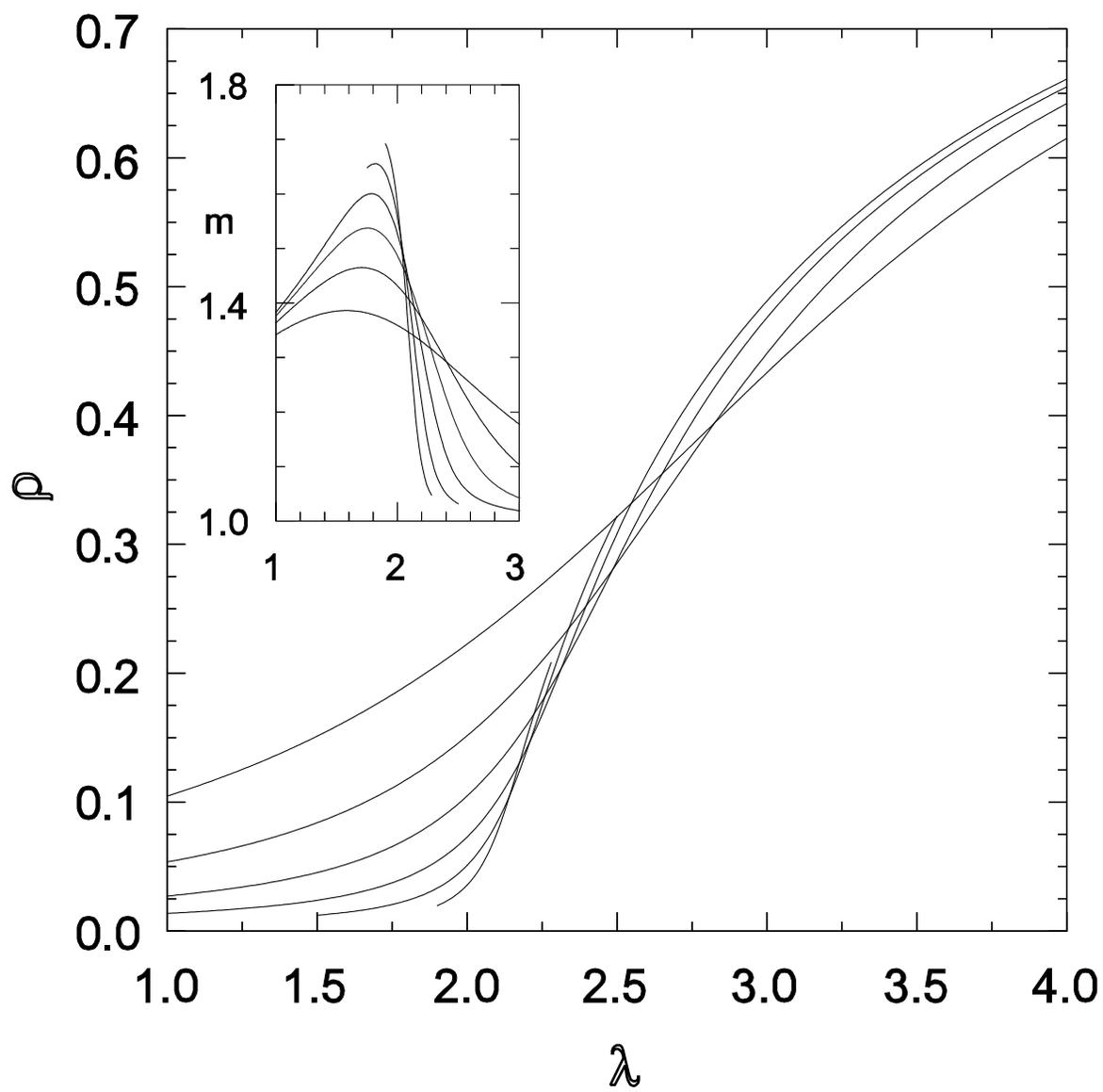
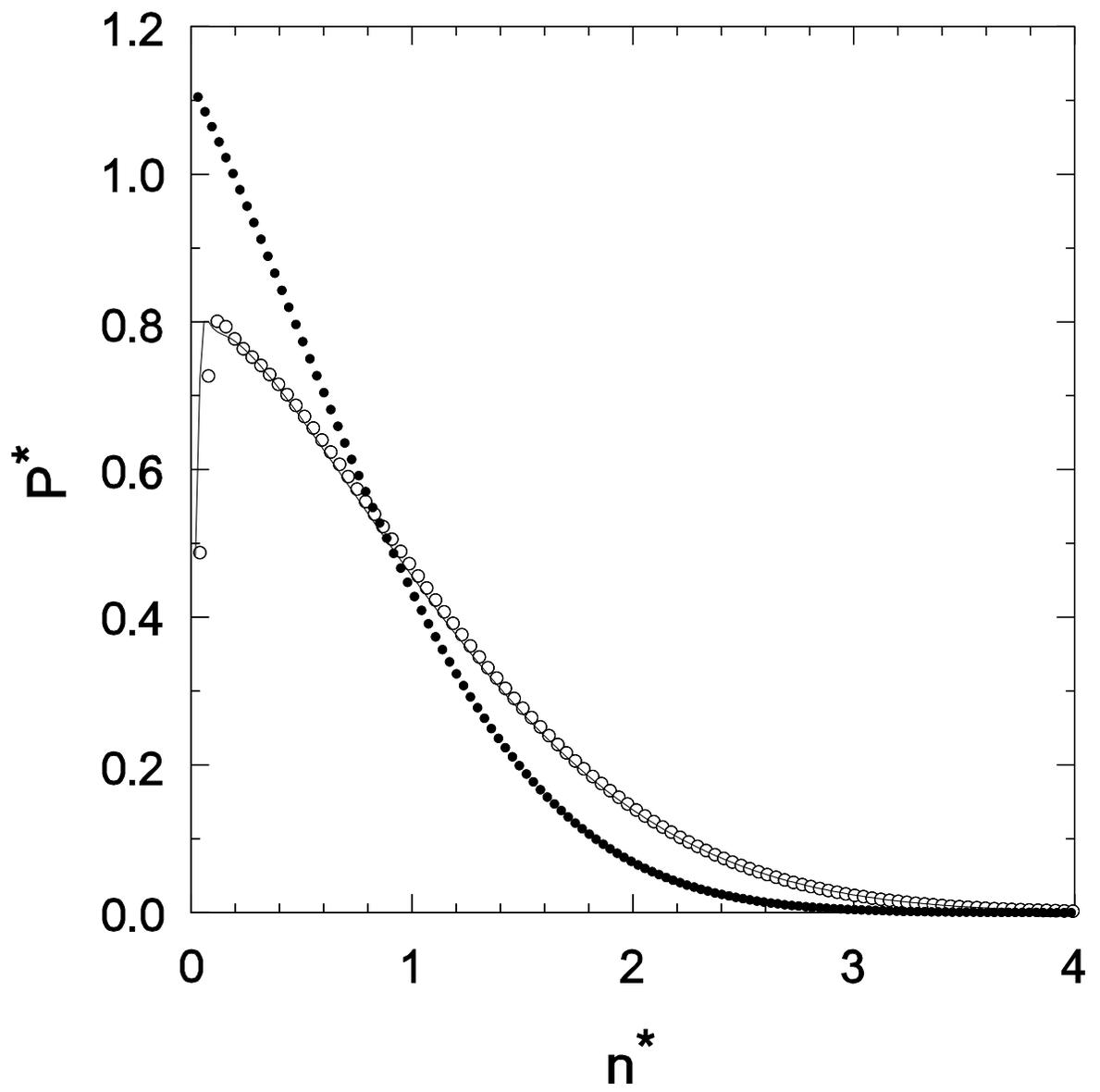
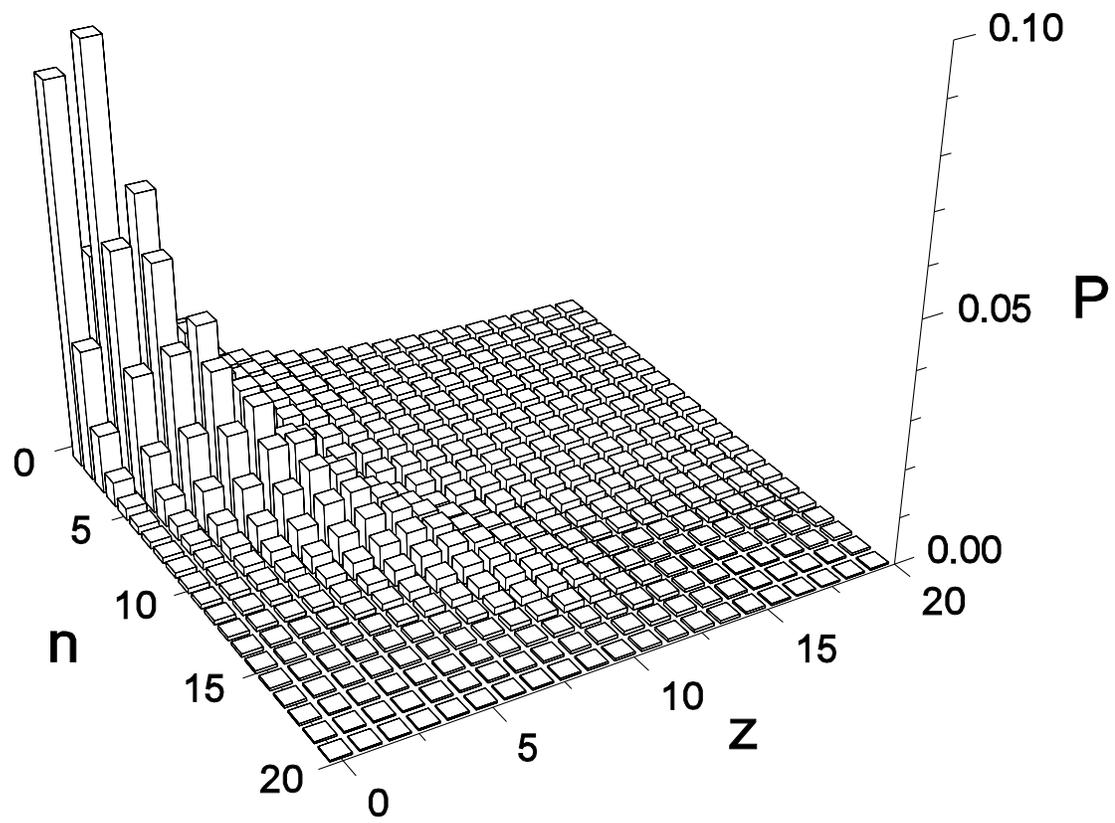
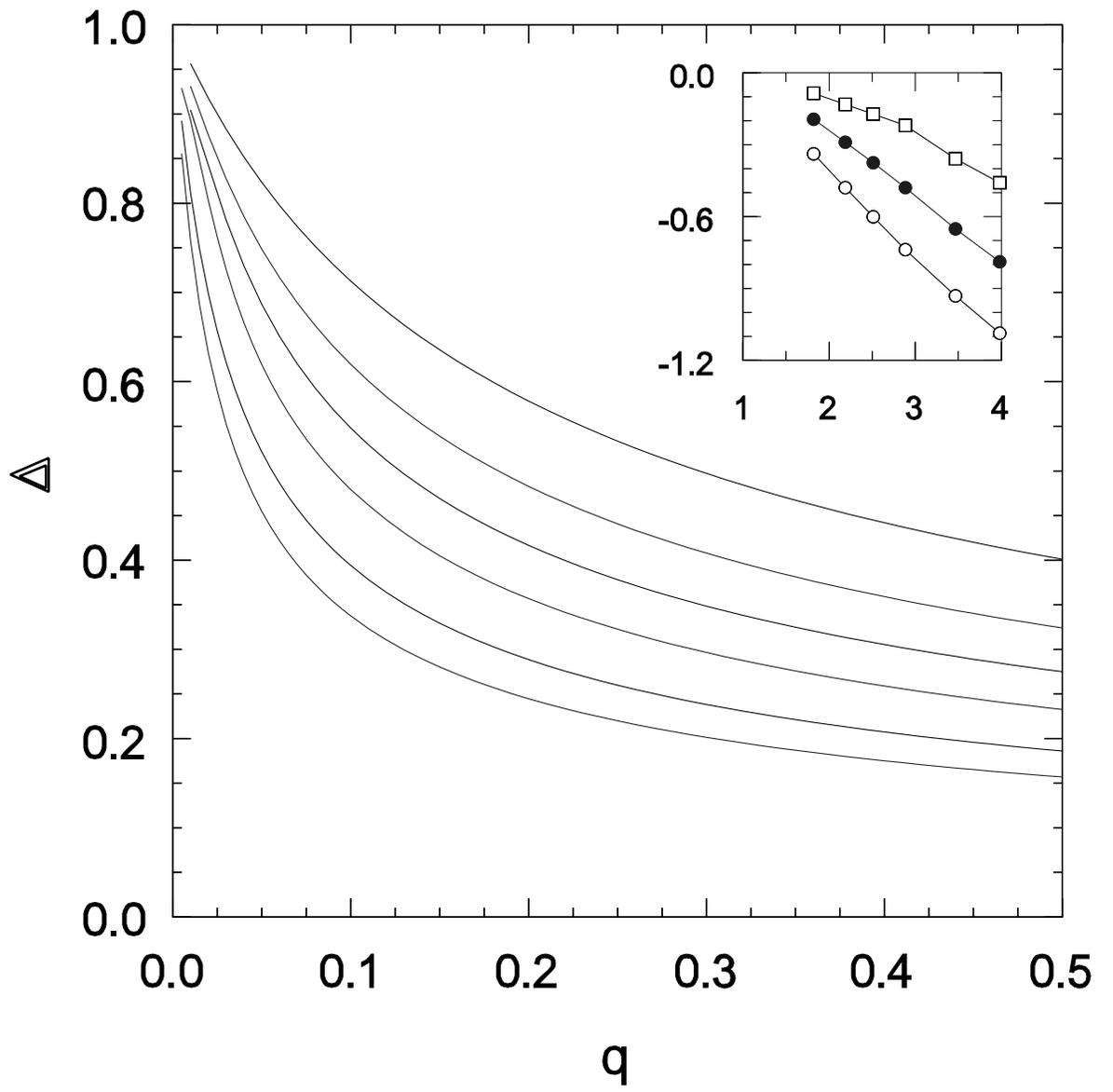
\end{document}